\documentclass[journal,draftcls,onecolumn,11pt]{IEEEtran}
\usepackage{amsmath,amssymb,graphicx,amsfonts,epsfig,cite,subfigure,times,latexsym,array,verbatim,url}

\newcommand{\emb}[1]{\text{\boldmath{$#1$}}}
\newcommand{\Prob}{\textsf{P}}
\newcommand{\Expect}{\textsf{E}}
\newcommand{\indic}[1]{\mbox{$1\!\!1$}{\{#1\}}}

\newcommand{\QMDP}{$\text{Q}_{\text{MDP}}$}
\newcommand{\term}{\mathcal{T}}

\graphicspath{{figures/}}

\title{Sensor Management for Tracking in Sensor Networks\thanks{This work was funded in part by a grant from the Motorola corporation, a U.S. Army Research Office MURI grant W911NF-06-1-0094 through a subcontract from Brown University at the University of Illinois, a NSF Graduate Research Fellowship, and by a Vodafone Fellowship.}}
\author{Jason~A.~Fuemmeler,~\IEEEmembership{Member,~IEEE}, George~K.~Atia,~\IEEEmembership{Member,~IEEE}, and Venugopal~V.~Veeravalli,~\IEEEmembership{Fellow,~IEEE}
\thanks{This work was done at the Coordinated Science Laboratory (CSL), University of Illinois at Urbana-Champaign, Urbana IL 61801, Emails: \{fuemmele,atia1,vvv\}@illinois.edu}}
\date{}

\begin{document}
\maketitle
\begin{abstract}
We study the problem of tracking an object moving through a network of wireless sensors. In order to conserve energy, the sensors may be put into a sleep mode with a timer that determines their sleep duration. It is assumed that an asleep sensor cannot be communicated with or woken up, and hence the sleep duration needs to be determined at the time the sensor goes to sleep based on all the information available to the sensor. Having sleeping sensors in the network could result in degraded tracking performance, therefore, there is a tradeoff between energy usage and tracking performance. We design sleeping policies that attempt to optimize this tradeoff and characterize their performance.  As an extension to our previous work in this area \cite{fuemmeler08}, we consider generalized models for object movement, object sensing, and tracking cost. For discrete state spaces and continuous Gaussian observations, we derive a lower bound on the optimal energy-tracking tradeoff. It is shown that in the low tracking error regime, the generated policies approach the derived lower bound. %Due to an inherent all-awake assumption, the lower bound is known to be loose in the high tracking error regime when the energy cost per sensor is too high. This fact, combined with the small gap at low tracking error, highlight the efficiency of our sensor sleeping policies.
\end{abstract}

\section{Introduction}
Large sensor networks collecting data in dynamic environments are typically composed of a distributed collection of cheap nodes with limited energy and processing capabilities. Hence, it is imperative to efficiently manage the sensors' resources to prolong the lifetime of such networks without sacrificing performance. Our focus in this paper is on sensor resource management for tracking and surveillance applications.

Previous work on sensor resource management considered the design of sensor sleeping protocols for sensor sleeping via wakeup mechanisms \cite{brooks03,balasubramanian04,gupta03,yang03,xu04,yang06} or by modifying power-save functions in MAC protocols for wireless ad hoc networks \cite{gui04,gui05,vasanthi06}. In the context of target classification, Castanon \cite{castanon} developed an approximate dynamic programming approach for dynamic scheduling of multi-mode sensors subject to sensors resource constraints. In \cite{asilomar_scheduling,tsp_scheduling} we studied a single object tracking problem where the sensors can be turned on or off at consecutive time steps to conserve energy (sensor scheduling). A controller selects the subset of sensors to activate at each time step. Also in \cite{fuemmeler08}, we studied a tracking problem where each sensor could enter a sleep mode with a sleep timer (sensor sleeping). While in sleep mode, the sensor could not assist in tracking the object by making observations. In contrast to \cite{tsp_scheduling}, in \cite{fuemmeler08} we assumed that sleeping sensors could not be woken up externally but instead had to set internal timers to determine the next time to come awake, wherefore, the control actions correspond to the sleep durations of awake sensors. In turn, this did not only entail a different control space, but also led to a significantly different policy design problem since a decision to put a sensor to sleep implies that this sensor cannot be scheduled at future time steps until it comes awake. The consequences of the current action on the tracking performance could be more dramatic rendering future planning more crucial. This led to a design problem that sought to optimize a tradeoff between energy efficiency and tracking performance. While optimal solutions to this problem could not be found, suboptimal solutions were devised that were demonstrated to be near optimal. To aid analysis, we assumed particularly simple models for object movement, object sensing, and tracking cost.  In particular, we assumed that the network could be divided into cells, each of which contained a single sensor.  The object moved among the cells and could only be observed by the sensor in the currently occupied cell.  Tracking performance was a binary quantity; either the object was observed in a particular time slot or it was not observed depending on whether the right sensor was awake.

In this paper, we continue to examine the fundamental theory of sleeping in sensor networks for tracking but we extend our analysis to more generalized models for object movement, object sensing, and tracking cost. We allow the number of possible object locations to be different from the number of sensors. The number of possible object locations can even be infinite to model the movement of an object on a continuum. Moreover, the object sensing model allows for an arbitrary distribution for the observations given the current object location, and the tracking cost is modeled via an arbitrary distance measure between the actual and estimated object location.

Not surprisingly, this generalization results in a problem that is much more difficult to analyze. Our approach is to build on the policies designed in \cite{fuemmeler08}. The design of those policies relied on the separation of the problem into a set of simpler subproblems. In \cite{fuemmeler08}, we have shown that under an observable-after-control assumption, the design problem lends itself to a natural decomposition into simpler per-sensor subproblems due to the simplified nature of the tracking cost structure. Unfortunately, this does not extend to the generalized cases we consider herein. However, based on the intuition gained from the structure of the solution in the simplified case, in this work we artificially separate our problem into a set of simpler per-sensor subproblems. The parameters of these subproblems are not known {\em a priori} due to the difficulties in analysis. However, we use Monte Carlo simulation and learning algorithms to compute these parameters. We characterize the performance of the resulting sleeping policies through simulation. For the special case of a discrete state space with continuous Gaussian observations, we derive a lower bound on the optimal energy-tracking tradeoff which is shown to be loose at the high tracking error regime, but is reasonably tight for the low tracking error region.

The remainder of this paper is organized as follows. In Section~\ref{sec:probform}, we describe the tracking problem in mathematical terms and define the optimization problem.  In Section~\ref{sec:subopt_sol} we derive our suboptimal solutions and the aforementioned lower bound. In Section~\ref{sec:num_res}, we provide numerical results that illustrate the efficacy of the proposed sleeping policies. We summarize and conclude in Section~\ref{sec:concl}.

\section{Problem Formulation} \label{sec:probform}
\subsection{POMDP Formulation}
Consider a network with $n$ sensors. Each sensor can be in one of two states: awake or asleep. A sensor in the awake state consumes more energy than one in the asleep state. However, object sensing can be performed only in the awake state.
We denote the set of possible object locations as $\mathcal{B}$ such that $|\mathcal{B}| = m+1$ where the $(m+1)$-th state represents an absorbing terminal state that occurs when the object leaves the network.  We also refer to this terminal state as $\term$.  If $\mathcal{B}$ is not a finite set then $m$ is $\infty$.  We define a {\em kernel} $P$ such that $P(x,\mathcal{Y})$ is the probability that the next object location is in the set $\mathcal{Y} \subset \mathcal{B}$ given that the current object location is $x$.  We can predict $t$ time steps into the future by defining $P^1 = P$ and $P^t$ inductively as
\begin{align}
   P^t(x,\mathcal{Y}) = \int_{\mathcal{B}} P^{t-1}(x,dz) P(z,\mathcal{Y})
\end{align}

Suppose $p$ is a probability measure on $\mathcal{B}$ such that $p(\mathcal{X})$ for $\mathcal{X} \in \mathcal{B}$ is the probability that the state is in $\mathcal{X}$ at the current time step.  Then the probability that the state will be in $\mathcal{Y}$ after $t$ time steps in the future is given by
\begin{align}
   (p P^t)(\mathcal{Y}) \equiv \int_{\mathcal{B}} p(dx) P^t(x,\mathcal{Y})
\end{align}
This defines the measure $p P^t$ which depends on both the prior $p$ and the transition Kernel $P$.  Let $b_k$ denote the state for the object at time $k$.  Also, let $\delta_x$ denote a probability measure such that $\delta_{x}(\mathcal{A}) = 1$ if $x \in \mathcal{A}$, and $\delta_{x}(\mathcal{A}) = 0$ otherwise. Conditioned on the object state $b_k$, the future state $b_{k+1}$ has a distribution $\delta_{b_k} P$. This defines the evolution of the object location. For a discrete state space this is simply the probability mass function defined by the $b_k$-th row of a transition matrix $P$.
%\begin{align} \label{eq:bkh}
%   b_{k+1} \sim \delta_{b_k} P
%\end{align}
We assume that it is always possible to determine if the object has left the network, i.e., if $b_k = m+1$.  To this end, we define a virtual sensor $n+1$ that detects without error whether the object has left the network. In other words, sensor $n+1$ is always awake but consumes no energy.

To provide a means for centralized control, we assume the presence of an extra node called the central controller.  The central controller keeps track of the state of the network and assigns sleep times to sensors that are awake.  In particular, each sensor that wakes up remains awake for one time unit during which the following actions are taken: (i) the sensor sends its observation of the object to the central unit, and (ii) the sensor receives a new sleep time (which may equal zero) from the central controller.  The sleep time input is used to initialize a timer at the sensor that is decremented by one time unit each time step.  When this timer expires, the sensor wakes up.  Since we assume that wakeup signals are impractical, this timer expiration is the only mechanism for waking a sensor.

Let $r_{k,\ell}$ denote the value of the sleep timer of sensor $\ell$ at time $k$.  We call the $(n+1)$-vector $\emb
{r}_k$ the residual sleep times of the sensors at time $k$.  Also, let $u_{k,\ell}$ denote the sleep time input supplied to sensor $\ell$ at time $k$.  We add the constraints $r_{k,n+1}=0$ and $u_{k,n+1}=0$ due to the nature of the virtual sensor $n+1$.  We can describe the evolution of the residual sleep times as
\begin{align} \label{eq:rklh}
   r_{k+1,\ell} = (r_{k,\ell} - 1) \indic{r_{k,\ell}> 0} + u_{k,\ell} \indic{r_{k,\ell} = 0}
\end{align}
for all $k$ and $\ell \in \{1,\dots,n+1\}$.  The first term on the right hand side of this equation expresses that if the sensor is currently asleep (the sleep timer for the sensor is not zero), the sleep timer is decremented by 1.  The second term expresses that if the sensor is currently awake (the sleep timer is zero), the sleep timer is reset to the current sleep time input for that sensor.

Based on the probabilistic evolution of the object location and \eqref{eq:rklh}, we see that we have a discrete-time dynamical model that describes our system with a well-defined state evolution.  The {\em state} of the system at time $k$ is described by $x_k = (b_k, \emb{r}_k)$. %and the state evolution is defined in \eqref{eq:bkh} and \eqref{eq:rklh}.
Unfortunately, not all of $x_k$ is known to the central unit at time $k$ since $b_k$ is known only if the object location is being tracked precisely.  Thus we have a dynamical system with incomplete (or partially observed) state information.

We write the observations for our problem as
\begin{align} \label{eq:zkh}
   z_k = (\emb{s}_k,\emb{r}_k)
\end{align}
where $\emb{s}_k$ is an $(n+1)$-vector of observations.  These observations are drawn from a probability measure $\sigma_{\emb{x}_k}$ that depends on $x_k$.
% We write this as
%\begin{align} \label{eq:skh}
%   \emb{s}_k \sim \sigma_{\emb{x}_k}
%\end{align}
However, we add two restrictions.  The first is that if a sensor is not awake at time $k$, its observation is an erasure.  Mathematically, we say that $r_{k,\ell} > 0$ implies $s_{k,\ell} = \mathcal{E}$.  The second restriction is that $s_{k,n+1}$ is a binary observation that indicates whether the object has left the network.

The total information available to the control unit at time $k$ is given by
\begin{align} \label{eq:Ikh}
   I_k = (z_0, \ldots, z_k, \emb{u}_0, \ldots, \emb{u}_{k-1})
\end{align}
with $I_0 = z_0$ denoting the initial (known) state of the system.
The control input for sensor $\ell$ at time $k$ is allowed to be a function of the information state $I_k$, i.e.,
\begin{align} \label{eq:mukh}
   \emb{u}_k = \mu_k(I_k)
\end{align}
The vector-valued function $\mu_k$ is the sleeping policy at time $k$ which defines a mapping from the information state $I_k$ to the set of admissible actions $\emb{u}_k$.

We now identify the costs present in our tracking problem.  The first is an {\em energy cost} of $c > 0$ for each sensor that is awake.  The energy cost can be written mathematically as
\begin{align}
   \sum_{\ell=1}^n c \indic{r_{k,\ell} = 0}
\end{align}
The second cost is a {\em tracking cost}.  To define the tracking cost, we first define the estimated object location at time $k$ to be $\hat{b}_k$.  We can think of $\hat{b}_k$ as an additional control input that is a function of $I_k$, i.e.,
\begin{align} \label{eq:betakh}
   \hat{b}_k = \beta_k(I_k)
\end{align}
Since $\hat{b}_k$ does not affect the state evolution, we do not need past values of this control input in $I_k$.  The tracking cost is a distance measure that is a function of the actual and estimated object locations and is written as
\begin{align}
   d(b_k,\hat{b}_k)
\end{align}
We assume that $d$ is a bounded function on $\mathcal{B} \times \mathcal{B}$.  Two examples of distance measures we might employ are the Hamming cost (if the space $\mathcal{B}$ is finite), i.e.,
\begin{align}
   d(b_k,\hat{b}_k) = \indic{\hat{b}_k \neq b_k}
\end{align}
and the squared Euclidean distance (if the space $\mathcal{B}$ is a subset of an appropriate vector space), i.e.,
\begin{align}
   d(b_k,\hat{b}_k) = \| \hat{b}_k - b_k \|_2^2
\end{align}
The parameter $c$ is used to trade off energy consumption and tracking errors.

Recall that the input $\hat{b}_k$ does not affect the state evolution; it only affects the cost. Therefore, we can compute the optimal choice of $\hat{b}_k$, given by $\beta_k^*(I_k)$, using an optimization minimizing the tracking error over a single time step.  We can thus write
\begin{align}
   \beta_k^*(I_k) = \arg \min_{\hat{b}} \Expect\left[
      d(b_k,\hat{b}_k)
   \Big|I_k \right]
\end{align}

Remembering that once the terminal state is reached no further cost is incurred, we can write the total cost for time step $k$ as
\begin{align} \label{eq:gh}
   g(b_k,I_k) = \indic{b_k  \neq \term} \left(
      d\left(b_k,\beta_k^*(I_k)\right)
      + \sum_{\ell=1}^n c \indic{r_{k,\ell} = 0}
   \right)
\end{align}
The infinite horizon cost for the system is given by
\begin{align} \label{eq:tot_costh}
   J(I_0, \mu_0, \mu_1, \ldots) = \Expect \left[ \left. \sum_{k=1}^{\infty} g(b_k,I_k) \right| I_0 \right]
\end{align}
Since $g$ is bounded (since the function $d$ is bounded) and the expected time till the object leaves the network is finite, the cost function $J$ is well defined.  The goal is to compute the solution to
\begin{align} \label{eq:opt_probh}
   J^*(I_0)=\min_{\mu_0, \mu_1, \hdots} J(I_0, \mu_0, \mu_1, \ldots)
\end{align}
The solution to this optimization problem for each value of $c$ yields an optimal sleeping policy.  The optimization problem falls under the framework of a partially observable Markov decision process (POMDP) \cite{aberdeen03,littman95,monahan,Hauskrecht2000}.

\subsection{Dealing With Partial Observability}
Partial observability presents a problem since the information for decision-making at time $k$ given in \eqref{eq:Ikh} is unbounded in memory.  To remedy this, we seek a sufficient statistic for optimization that is bounded in memory.  The observation $\emb{s}_k$ depends only on $x_k$, which in turn depends only on $x_{k-1}$, $u_{k-1}$, and some random disturbance $w_{k-1}$.  It is a standard argument (e.g., see \cite{bertsekas07}) that for such an observation model, a sufficient statistic is given by the probability distribution of the state $x_k$ given $I_k$.  Such a sufficient statistic is referred to as a {\em belief state} in the POMDP literature (e.g., see \cite{aberdeen03,littman95}).  Since the residual sleep times portion of our state is observable, the sufficient statistic can be written as $v_k = (p_k, \emb{r}_k)$, where $p_k$ is a probability measure on $\mathcal{B}$.  Mathematically, we have
\begin{align} \label{eq:pkldefh}
   p_k(\mathcal{X}) = \Prob(b_k \in \mathcal{X}|I_k)
\end{align}
The task of recursively computing $p_k$ for each $k$ is a problem in nonlinear filtering (e.g., see \cite{doucet01}).  In other words, $p_{k+1}$ can be computed using standard Bayesian techniques as the posterior measure resulting from prior measure $p P$ and observations $\emb{s}_{k+1}$.

The function $\beta_k^*$ that determines $\hat{b}_k$ can now be written in terms of $p_k$ and $\emb{r}_k$ instead of $I_k$.  We can rewrite it as
\begin{align}
   \beta_k^*(p_k,\emb{r}_k)  & = \arg \min_{\hat{b}} \Expect\left[
      d(b_k,\hat{b}) | b_k \sim p_k
   \right] \\
   & = \arg \min_{\hat{b}} \int_{\mathcal{B}} d(b_k,\hat{b}) p_k(db)
\end{align}
Note that due to the stationarity of the state evolution, $\beta_k^*$ has the same form for every $k$ and is independent of $\emb{r}_k$.  Thus, we can drop the subscript and refer to $\beta_k^*$ as $\beta^*$, a function of $p_k$ alone.

Now we write our dynamic programming problem in terms of the sufficient statistic.  We first rewrite the cost at time step $k$.  Since only expected values of the cost function $g$ appear in \eqref{eq:tot_costh}, we can take our cost function to be the expected value of $g$ (defined in \eqref{eq:gh}) conditioned on $b_k$ being distributed according to $p_k$. With a slight abuse of notation, we call this redefined cost $g$. The cost can then be written as
\begin{align}
   g(p_k,\emb{r}_k) & = \int_{\mathcal{B}} \indic{b  \neq \term} \left(
      d(b, \beta^*(p_k))
      + \sum_{\ell=1}^n c \indic{r_{k,\ell} = 0}
   \right) p_k(db) \\
   & = \int_{\mathcal{B} - \term} \left(
      d(b, \beta^*(p_k))
      + \sum_{\ell=1}^n c \indic{r_{k,\ell} = 0}
   \right) p_k(db)
\end{align}
The selection of sleep times, originally presented in \eqref{eq:mukh}, can now be rewritten as
\begin{align}
   \emb{u}_k = \mu_k(p_k,\emb{r}_k)
\end{align}
The total cost defined in \eqref{eq:tot_costh} becomes
\begin{align}
   J(p_0,\emb{r}_0, \mu_0, \mu_1, \ldots) = \Expect \left[ \left. \sum_{k=1}^{\infty} g(p_k,\emb{r}_k) \right| v_0 \right]
\end{align}
and the optimal cost defined in \eqref{eq:opt_probh} becomes
\begin{align}
   J^*(p_0,\emb{r}_0)=\min_{\mu_0,\mu_1, \hdots} J(p_0,\emb{r}_0,\mu_0, \mu_1, \hdots)
\end{align}

\section{Suboptimal Solutions} \label{sec:subopt_sol}

Similar to the problem in \cite{fuemmeler08}, an optimal policy could be found by solving the Bellman equation
\begin{align} \label{eq:bellmanh}
   J(p,\emb{r}) = \min_{\mu} \Expect \left[ g(p_1,\emb{r}_1) + J(p_1,\emb{r}_1) | p_0=p, \emb{r}_0 = \emb{r}, \emb{u}_0 = \mu(p_0,\emb{r}_0) \right]
\end{align}
However, since an optimal solution could not be found for the simpler problem considered in \cite{fuemmeler08}, we immediately turn our attention to finding suboptimal solutions to our problem.

Note that in \cite{fuemmeler08}, simpler sensing models and cost structures were employed. Under a simplifying observable-after-control assumption, the simplicity of the sensing models allowed for the decoupling of the contributions of the individual sensors.  The simplicity of the cost structures allowed the cost to be written as a sum of per-sensor costs.  The result was a problem that could be written as a number of simpler subproblems. The present case is more complicated.  In general, the cooperation among the sensors may be difficult to analyze and understand.  Furthermore, the tracking cost may not be easily written as a sum across the sensors.

Based on the intuition gained from \cite{fuemmeler08}, our approach to generating suboptimal solutions is to artificially write the problem as a set of subproblems that can be solved using the techniques of \cite{fuemmeler08}. The tracking cost expressions (which are a function of the sleeping actions of the sensors) in these subproblems will be left as unknowns. To determine appropriate values for these tracking costs, we either perform Monte Carlo simulations before tracking begins or use data gathered during tracking.  The intuition is that if the resultant tracking cost expressions capture the ``typical'' behavior of the actual tracking cost, then our sleeping policies should perform well.

\subsection{General approach}
The complexity of the sleeping problem stems from:
\begin{enumerate}
\item The complicated evolution of the belief state $p_k$ (non-linear filtering).
\item The complexity of the model including the dimensionality of the state space, the control space and the observation space.
\end{enumerate}
To address the aforementioned difficulties, our approach has two main ingredients. First, we make assumptions about the observations that will be available to the controller at future time steps. To generate sleeping policies, we assume that the system is either perfectly observable or totally unobservable after control. Hence, we define approximate recursions with special structure as surrogates for the optimal value function. Second, we devise different methodologies to evaluate suitable tracking costs in Sections \ref{sec:non_learning} and \ref{sec:learning} whereby we capture the effect of each sensor on the overall tracking cost. Writing the combined tracking cost as the sum of independent contributions of different sensors (with respect to some baseline) allows us to write the Bellman equation as the sum of per-sensor recursions. Instead of solving the Bellman equation in (\ref{eq:bellmanh}), we alternatively solve $n$ simpler Bellman equations to find per-sensor policies and cost functions. The overall policy is then the per-sensor policies applied in parallel.

We denote by $J^{(\ell)}$ the cost function of the $\ell$-th sensor approximate subproblem. We define $T^\Delta(b,\ell)$ to be the increase in tracking cost due to not waking up sensor $\ell$ at time $k$ given that $b_{k-1} = b$. This is meant to capture the contribution of the $\ell$-th sensor to the total tracking cost. Next we define our approximations.
%As we show, this addresses the first difficulty associated with the belief evolution.

\subsubsection{\QMDP{}}
First introduced in the artificial intelligence literature \cite{cassandra97,QMDP}, the \QMDP{} solution for POMDPs assumes that the system will be perfectly observable after control, i.e., the partially observable state becomes fully observable after taking a control action. In other words, under a \QMDP{} assumption the belief state simply evolves as
\begin{align}
p_{k+1}=\delta_{b_{k+1}}
\label{eq:QMDP_belief_evolution}
\end{align}
Noting that the future cost is not only affected by the current control action through belief evolution, but also by the fact that no future decisions can be made for a sleeping sensor until it wakes up, the observable-after-control policy is by no means a myopic policy. Note that (\ref{eq:QMDP_belief_evolution}) does not imply zero tracking errors; it is merely an assumption simplifying the state evolution in order to generate a sleeping policy. Now we can readily define a {\em \QMDP{} per-sensor Bellman equation} analogous to the one in \cite{fuemmeler08} as
\begin{align} \label{eq:qmdpbellh}
   J^{(\ell)}(p) = \min_{u} \left(
      \sum_{j=0}^{u-1} \int_{\mathcal{B}-\term} T^\Delta(b,\ell) \, (p P^j)(db) +
      \int_{\mathcal{B}-\term} \left(
         c + J^{(\ell)}(\delta_b)
      \right)  \, (p P^{u+1})(db)
   \right)
\end{align}
To clarify, the first summation in the R.H.S. of (\ref{eq:qmdpbellh}) corresponds to the expected tracking cost incurred by the sleep duration $u$ of sensor $\ell$. The second term consists of: (i) the energy cost incurred as the sensor comes awake after its sleep timer expires (after $u+1$ time slots); and (ii) the cost to go under an observable-after-control assumption (hence the belief state is $\delta_b$).

We cannot find an analytical solution for (\ref{eq:qmdpbellh}). However, note that if we can solve (\ref{eq:qmdpbellh}) for $p=\delta_b$ for all $b$, then it is straightforward to find the solution for all values of $p$. Thus, given a function $T^\Delta$, \eqref{eq:qmdpbellh} can be solved through standard policy iteration \cite{bertsekas07}, but only if $\mathcal{B}$ is finite.

\subsubsection{FCR}
Similarly, we define a First Cost Reduction {\em (FCR) Bellman equation} analogous to the one in \cite{fuemmeler08} as
\begin{align} \label{eq:fcrbellh}
   J^{(\ell)}(p) = \min_{u} \left(
      \sum_{j=0}^{u-1} \int_{\mathcal{B}-\term} T^\Delta(b,\ell) \, (p P^j)(db) +
      c \int_{\mathcal{B}-\term}  (p P^{u+1})(db) +
      J^{(\ell)}(p P^{u+1})
   \right)
\end{align}
In this case, it is assumed that we will have no future observations. In other words, we define the belief evolution as $p_{k+1}=p_kP$. Again, it is worth mentioning that this does not mean that it would be impossible to track the object; we are simply making a simplifying assumption about the future state evolution in order to generate a sleeping policy. Given a function $T^\Delta$, it is easy to verify that the solution to \eqref{eq:fcrbellh} is
\begin{align}
   J^{(\ell)}(p) = \sum_{j=0}^{\infty} \min \left\{
      \int_{\mathcal{B}-\term} T^\Delta(b,\ell) \, (p P^j)(db),
      c \int_{\mathcal{B}-\term}  (p P^{j+1})(db)
   \right\}
\end{align}
and the associated policy is to choose the first value of $u$ such that
\begin{align}
      c \int_{\mathcal{B}-\term}  (p P^{u+1})(db) \geq
      \int_{\mathcal{B}-\term} T^\Delta(b,\ell) \, (p P^u)(db),
\end{align}
In other words, the policy is to come awake at the first time the expected tracking cost exceeds the expected energy cost where the tracking cost is defined based on $T^\Delta$ (to be determined) hence the name First Cost Reduction.

The solutions to the per-sensor Bellman equations in (\ref{eq:qmdpbellh}) and (\ref{eq:fcrbellh}) define the \QMDP{} and FCR policies for each sensor, respectively. Note that, unlike \cite{fuemmeler08,asilomar_scheduling,tsp_scheduling}, the solution to the \QMDP{} recursion does not necessarily provide a lower bound on the optimal value function since the employed tracking cost is not a lower bound on the actual tracking cost. In Sec \ref{sec:lower_bound} we derive a lower bound on the optimal energy-tracking tradeoff for discrete state spaces with Gaussian Observations. The remaining task is to identify appropriate values of $T^\Delta(b,\ell)$ for all $b \neq \term$ and for all $\ell$.  This is the subject of the next two sections.

\subsection{Nonlearning approach}
\label{sec:non_learning}
For now, suppose that $\mathcal{B}$ is a finite space.  Suppose $b_{k-1} = b$.  To generate $T^\Delta(b,\ell)$ for a particular $\ell$, we first assume a ``baseline'' behavior for the sensors, i.e., we make an assumption about the set of sensors that are awake at time $k$ given that $b_{k-1} = b$.  We consider two possibilities:
\begin{enumerate}
   \item That all sensors are asleep.
   \item That the set of sensors awake is selected through a greedy algorithm.  In other words, the sensor that causes the largest decrease in expected tracking cost is added to the awake set until any further reduction due to a single sensor is less than $c$.  The expected tracking cost can be evaluated through the use of Monte Carlo simulation (repeatedly simulating our system from time $k-1$ to time $k$) to avoid the need for numerical integration.
\end{enumerate}
Starting with this set of awake sensors, the value of $T^\Delta(b,\ell)$ is then computed as the absolute difference in expected tracking cost incurred by changing the state of sensor $\ell$.  Again, Monte Carlo simulation can be used to evaluate the change in expected tracking cost.  We can think of this procedure as linearizing the tracking cost about some baseline behavior.

If $\mathcal{B}$ is not finite, then a parameterized version of $T^\Delta$ can be computed instead.  We choose $\tilde{m}$ elements of $\mathcal{B} - \term$ and evaluate $T^\Delta$ at these points.  The value of $T^\Delta$ at all other values of $b \in \mathcal{B} - \term$ can be computed via an interpolation algorithm.  Recall that only an FCR policy is appropriate in the infinite state case, since solving the \QMDP{} Bellman equation for an infinite number of point mass distributions is infeasible.

\subsection{Learning approach}  \label{sec:learning}
In this section, we describe an alternative learning-based approach. For ease of exposition, suppose that $\mathcal{B}$ is a finite space.  Then our probability measure $p_k$ can be characterized by a probability mass function.  We refer to this probability mass function as $\emb{p}_k$ (a row vector).  Define $\hat{a}_{k,\ell}$ to be the approximated expected increase in tracking cost due to sensor $\ell$ sleeping at time $k$ as
\begin{align}
   \hat{a}_{k,\ell} = \sum_{b \neq \term} \emb{p}_{k-1}(b) T^\Delta(b,\ell)
\end{align}
Ideally, we would like this approximation to be equal to the actual expected increase in tracking cost due to sensor $\ell$ sleeping.  Unfortunately, we do not have access to actual tracking costs at time $k$ since $b_k$ is not known exactly.  However, we do have access to $\emb{p}_k$, $\emb{r}_k$, and $\emb{p}_{k-1}$.  It is therefore possible to estimate the tracking cost as
\begin{align}
   \int_{\mathcal{B}} d(b,\beta^*(\emb{p}_k)) \emb{p}_k(db)
\end{align}
For example, if Hamming cost is being used, then we can estimate the tracking cost as
\begin{align}
   1 - \max_{b} p_k(\{b\})
\end{align}
and if squared Euclidean distance is being used we can estimate the tracking cost using the variance of the measure $p_k$. Next we describe how we learn $T^\Delta$ by solving a least squares problem.

Determining an estimate of the {\em increase} in the tracking cost due to the sleeping of sensor $\ell$ at time $k$, denoted $a_{k,\ell}$, depends on the value of $r_{k,\ell}$.  If $r_{k,\ell} = 0$, we ignore the observation from sensor $\ell$ and generate a new version of $\emb{p}_k$ called $\emb{p}^\prime_k$.  We can compute $a_{k,\ell}$ as
\begin{align}
   a_{k,\ell} = \sum_{b \neq \term} \emb{p}^\prime_k(b) d(b,\beta^*(\emb{p}^\prime_k)) - \sum_{b \neq \term} \emb{p}_k(b) d(b,\beta^*(\emb{p}_k))
\end{align}
If on the other hand $r_{k,\ell} > 0$, we we first generate an object location $b^\prime_k$ according to $\emb{p}_k$ and then generate an observation according to the probability measure $\sigma_{b^\prime_k}$.  This observation is used to generate a new distribution $\emb{p}^\prime_k$ from $\emb{p}_k$.  Then we compute $a_{k,\ell}$ as
\begin{align}
   a_{k,\ell} = \sum_{b \neq \term} \emb{p}_k(b) d(b,\beta^*(\emb{p}_k)) - \sum_{b \neq \term} \emb{p}^\prime_k(b) d(b,\beta^*(\emb{p}^\prime_k))
\end{align}

We now have an approximation sequence $\hat{a}_{k,\ell}$ and an observation sequence $a_{k,\ell}$. At time $k-1$, our goal is to choose $T^\Delta$ to minimize
\begin{align}
   \Expect\left[ ( \hat{a}_{k,\ell} - a_{k,\ell} )^2 \right]
\end{align}
We apply the Robbins-Monro algorithm, a form of stochastic gradient descent, to this problem in order to recursively compute a sequence of $T^\Delta$ that will hopefully solve this minimization problem for large $k$.  The update equation is
\begin{align}
   T^\Delta_k(b,\ell) = T^\Delta_{k-1}(b,\ell) - 2 \alpha_k \indic{b \neq \term} \emb{p}_{k-1}(b) (\hat{a}_{k,\ell} - a_{k,\ell} )
\end{align}
where $\alpha_k$ is a step size.  Note that $\indic{b \neq \term} \emb{p}_{k-1}(b)$ is the gradient of $\hat{a}_{k,\ell}$ with respect to $T^\Delta(b,\ell)$.

Using a constant step size in our simulations, we could only observe small oscillations in the values of $T^\Delta$. It is unclear whether there are conditions under which the local or global convergence of this learning algorithm is guaranteed.  The difficulty is that the observations we are trying to model depend on the model itself.  The problem is reminiscent of optimistic policy iteration (see \cite{bertsekas07}), the convergence properties of which are little understood.  We have left a proof of convergence for future work.  It should be pointed out that the algorithm will likely converge more slowly for a two-dimensional network than a one-dimensional network.  The reason is that in two dimensions it is easier for an object to avoid visiting an object location state and causing an update to that particular value of $T^\Delta$.

If $\mathcal{B}$ is not finite, then we can again parameterize $T^\Delta$ as in the previous section.  The Robbins-Monro algorithm can be applied in this context as well, although the gradient expressions will depend on the type of interpolation used.

\subsection{A Lower Bound}
\label{sec:lower_bound}
Unfortunately, deriving a lower bound is generally difficult for the considered problem. However, in this section we derive a lower bound for the special case of a discrete state space with Gaussian observations. Our approach is similar to \cite{tsp_scheduling} in which we considered a related scheduling problem. The idea is to combine the observable-after-control assumption with a separable lower bound on the tracking cost as we demonstrate in what follows.

Given the current belief $\emb{p}_k$, an action vector $\emb{u}_k$, and the current residual sleep times vector $\emb{r}_k$, the expected tracking cost can be written as:
\begin{eqnarray}
E[d(\hat{b}_{k+1},b_{k+1})|\emb{p}_k,\emb{u}_k,\emb{r}_k]&=& \sum_{j=1}^m \Pr[\hat{b}_{k+1}\ne j|p_k,\emb{u}_k, \emb{r}_k, b_{k+1}=j]\Pr[b_{k+1}=j|\emb{p}_k,\emb{u}_k]\nonumber\\
&=& \sum_{i=1}^m \emb{p}_k(i)\sum_{j=1}^m p(b_{k+1}=j|b_k=i)\Pr[\hat{b}_{k+1}\ne j|\emb{p}_k,\emb{u}_k,\emb{r}_k,b_{k+1}=j]\nonumber\\
\end{eqnarray}

When awake, the sensors observations are Gaussian, i.e.,
\begin{align}\label{eq:gaussian_obs}
   s_{k,\ell} \sim \mathcal{N}\left(
      \frac{10}{(\nu_\ell - b_k)^2 + 1},1
   \right)
\end{align}
where $\nu_\ell$ is the location of sensor $\ell$.

Defining,
\[
P(E|H_j)\triangleq \Pr[\hat{b}_{k+1}\ne j|\emb{p}_k,\emb{u}_k,\emb{r}_k,b_{k+1}=j]
\]
which is a conditional error probability for a multiple hypothesis testing problem with $m$ hypotheses, each corresponding to a different mean vector contaminated with white Gaussian noise. Conditioned on $H_j$, the observation model is:
\begin{equation}
H_j: \emb{s}(\ell) = (m_j(\ell) + w)\indic{r_{k+1,\ell}=0} + \varepsilon \indic{r_{k+1,\ell}>0}
\label{eq,underHj}
\end{equation}
where $\emb{s}(\ell)$ is the $\ell$-th entry of an $n\times 1$ vector $\emb{s}$ denoting the received signal strength at the $n$ sensors, $m_j$ is the mean received signal strength when the target is at state $j$ ($j$-th hypothesis) and $w$ is a zero mean white Gaussian Noise, i.e. $w\sim {\cal N}(0,\sigma^2)$. According to (\ref{eq,underHj}), if awake at the next time step, sensor $\ell$ gets a Gaussian observation that depends on the future target location, and an erasure, otherwise. Since the current belief is $\emb{p}_k$, the prior for the $j$-th hypothesis is $\pi_j=[\emb{p}_k P]_j$.

The error event $E$ can be written as the union of pairwise error regions as
\begin{eqnarray}
p(E|H_j) = \Pr[\cup_{k\ne j}\zeta_{kj}]
\end{eqnarray}
where \[\zeta_{kj}= \{\emb{s}:L_{kj}(\emb{s})>\frac{\pi_j}{\pi_k}\}\]
is the region of observations for which the $k$-th hypothesis $H_k$ is more likely than the $j$-th hypothesis $H_j$, and where
\[
L_{kj}\triangleq \frac{f(\emb{s}|H_k)}{f(\emb{s}|H_j)}
\]
denotes the likelihood ratio for $H_k$ and $H_j$.

Using standard analysis for likelihood ratio tests ~\cite{poor,levy}, it is not hard to show that:
\begin{equation}
 p(\zeta_{kj}|H_j)=Q\left(\frac{\emb{d}_{kj}}{2}+\frac{\ln\frac{\pi_j}{\pi_k}}{\emb{d}_{kj}}\right)
 \end{equation}
where $\emb{d}^2_{kj}=\frac{\emb{\Delta m}_{kj}^T\emb{\Delta m}_{kj}}{\sigma^2}$, $\emb{\Delta m}_{kj}=\emb{m}_k-\emb{m}_j$, and $Q(.)$ is the normal distribution $Q$-function. The quantity $\emb{d}_{kj}$ plays the role of distance between the two hypothesis and hence depends on the difference of their corresponding mean vectors and the noise variance $\sigma^2$. Hence, $\emb{d}_{kj}$ is a function of the next step residual sleep vector $\emb{r}_{k+1}$. To highlight this dependence, we will sometimes use the notation $\emb{d}_{kj}(\emb{r})$ when needed. Note that, for different values of $k$ and $j$, $\zeta_{kj}$ are not generally disjoint but allow us to lower bound the error probability in terms of pairwise error probabilities, namely, a lower bound can be written as:
\begin{eqnarray}
p(E|H_j)\geq\max_{k\ne j} p(\zeta_{kj}|H_j)
\end{eqnarray}
And we can readily lower bound the expected tracking error:
\begin{eqnarray}
E[d(\hat{b}_{k+1},b_{k+1})|\emb{p}_k,\emb{u}_k]&\geq&\sum_{i=1}^m \emb{p}_k(i)\sum_{j=1}^m p(b_{k+1}=j|b_k=i)\max_{k\ne j} p(\zeta_{kj}|H_j)\nonumber\\
& = &\sum_{i=1}^m \emb{p}_k(i)\sum_{j=1}^m p(b_{k+1}=j|b_k=i)\max_{k\ne j}Q\left(\frac{\emb{d}_{kj}}{2}+\frac{\ln\frac{\pi_j}{\pi_k}}{\emb{d}_{kj}}\right)
\label{eq:lb_tracking}
\end{eqnarray}

Next we separate out the effect of each sensor on the tracking error:
\begin{eqnarray}
E[d(\hat{b}_{k+1},b_{k+1})|\emb{p}_k,\emb{u}_k,\emb{r}_k]&\buildrel (a)\over\geq&\indic{r_{k+1,\ell}=0}E[d(\hat{b}_{k+1},b_{k+1})|\emb{p}_k,\emb{r}_{k+1}=\emb{0}]\nonumber\\
&+& \indic{r_{k+1,\ell}>0}E[d(\hat{b}_{k+1},b_{k+1})|\emb{p}_k,r_{k+1,i}=0 ~~\forall i\ne\ell]~~\mbox{for every}~~\ell\nonumber\\
\label{eq,trick}
\end{eqnarray}
where $\emb{0}$ is the all zero vector designating that all sensors will be awake at the next time slot. The inequality in
(a) follows from the fact that if we separate out the effect of the $\ell$-th sensor we get a better tracking performance when all the remaining sensors are awake. Since this holds for every $\ell$, a lower bound on the expected tracking error can be written as a convex combination of all sensors contributions:
\begin{align}
E[d(\hat{b}_{k+1},b_{k+1})|\emb{p}_k,\emb{u}_k,\emb{r}_k]\geq
\sum_{\ell=1}^n \lambda_{\ell}(\emb{p}_k)\Big\{&\indic{r_{k+1,\ell}=0}E[d(\hat{b}_{k+1},b_{k+1})|\emb{p}_k,\emb{r}_{k+1}=\emb{0}]\nonumber\\
&+ \indic{r_{k+1,\ell}>0}E[d(\hat{b}_{k+1},b_{k+1})|\emb{p}_k,r_{k+1,i}=0 ~~\forall i\ne\ell]\Big\}
\end{align}
where $\sum_{\ell}\lambda_{\ell}(\emb{p}_k) = 1$.

Let $\emb{0}_{-\ell}$ denote a vector of length $n$ with all entries equal to zero except for the $\ell$-th entry which can be anything greater than $0$. Then replacing from (\ref{eq:lb_tracking}),
\begin{align}
E[d(\hat{b}_{k+1},b_{k+1})|&\emb{p}_k,\emb{u}_k,\emb{r}_{k}]\geq\nonumber\\
\sum_{\ell=1}^n\lambda_{\ell}(\emb{p}_k)\Bigg\{&1{\hskip -2.5 pt}\hbox{I}_{\{r_{k+1,\ell}=0\}}\sum_{i=1}^m \emb{p}_k(i)\sum_{j=1}^m p(b_{k+1}=j|b_k=i)\max_{k\ne j}Q\left(\frac{\emb{d}_{kj}(\emb{0})}{2}+\frac{\ln\frac{\pi_j}{\pi_k}}{\emb{d}_{kj}(\emb{0})}\right)\nonumber\\
+&\indic{r_{k+1,\ell}>0}\sum_{i=1}^m \emb{p}_k(i)\sum_{j=1}^m p(b_{k+1}=j|b_k=i)\max_{k\ne j}Q\left(\frac{\emb{d}_{kj}(\emb{0}_{-\ell})}{2}+\frac{\ln\frac{\pi_j}{\pi_k}}{\emb{d}_{kj}(\emb{0}_{-\ell})}\right)\Bigg\}
\end{align}

To simplify notation, we introduce the following $2$ quantities:
\[T_0(\emb{p};i,\ell)\triangleq\sum_{j=1}^m p(b_{k+1}=j|b_k=i)\max_{k\ne j}Q\left(\frac{\emb{d}_{kj}(\emb{0})}{2}+\frac{\ln\frac{[\emb{p}P]_j}{[\emb{p}P]_k}}{\emb{d}_{kj}(\emb{0})}\right)\]

\[T(\emb{p};i,\ell) \triangleq\sum_{j=1}^m p(b_{k+1}=j|b_k=i)\max_{k\ne j}Q\left(\frac{\emb{d}_{kj}(\emb{0}_{-\ell})}{2}+\frac{\ln\frac{[\emb{p}P]_j}{[\emb{p}P]_k}}{\emb{d}_{kj}(\emb{0}_{-\ell})}\right)\]
Intuitively, $T_0(\emb{p};i,\ell)$ represents the contribution of sensor $\ell$ to the total expected tracking cost when the underlying state is $i$, the belief is $\emb{p}$ and when all sensors are awake. On the other hand $T(\emb{p};i,\ell)$ is the $\ell$-th sensor contribution when it is asleep and all the other sensors are awake.

Now if we assume that the target will be perfectly observable after taking the sleeping action, a lower bound on the total cost can be obtained from the solution of the following Bellman equation:
\begin{equation}
J(\emb{p},\emb{r}_0) = \sum_{\ell} J^{\ell}(\emb{p},r_{0,\ell})
\label{eq,decomposed_valfunc}
\end{equation}
where,
\begin{align}
J^{\ell}(\emb{p},r_{0,\ell})=\min_{u_{\ell}}\Bigg\{&\indic{r_{1,\ell}=0}\left(\sum_b \emb{p}(b)\lambda_{\ell}T_0(\emb{p};b,\ell)+c\sum_{i=1}^m [\emb{p}P]_i + \sum_{i=1}^m [\emb{p}P]_i J(\emb{e}_i,0)\right)\nonumber\\
&+\indic{r_{1,\ell}>0}\left(\sum_b \emb{p}(b)\lambda_{\ell}T(\emb{p};b,\ell)+\sum_{i=1}^m [\emb{p}P]_i J(\emb{e}_i,u_{\ell})\right)\Bigg\}
\label{eq:lb_valfunc_per_sensor}
\end{align}
Note that if we can solve the equation above for $\emb{p}=\emb{e}_i$ for all $i\in\{1,\ldots,m\}$, then it is straightforward to find the solution for all other values of $\emb{p}$. We therefore focus on specifying the value function at those points. Since this is the case, we further simplify our notation and use $T(i,\ell)$ and $\lambda(i,\ell)$ as shorthand for $T(\emb{e}_i;i,\ell)$ and $\lambda_{\ell}(\emb{e}_i)$, respectively. Also since an action only needs to be made when the sensor wakes up, we only need to define actions at $r_{0,\ell}=0$. Observing that
\begin{eqnarray}
J^{\ell}(\emb{e}_j,u)= \lambda(j,\ell)T(j,\ell)+ \sum_{i=1}^m [\emb{e}_jP]_i J^{\ell}(\emb{e}_i,u-1)~~~~~\forall~u>1
\label{eq,recursive_any_u}
\end{eqnarray}
and
\begin{eqnarray}
J^{\ell}(\emb{e}_j,1)= \lambda(j,\ell)T_0(j,\ell)+ c\sum_{i=1}^m [\emb{e}_jP]_i+\sum_{i=1}^m [\emb{e}_jP]_i J^{\ell}(\emb{e}_i,0)
\label{eq,recursive_u_equals_1}
\end{eqnarray}
we recursively substitute from (\ref{eq,recursive_any_u}) and (\ref{eq,recursive_u_equals_1}) in (\ref{eq:lb_valfunc_per_sensor}) until the system reaches $(\emb{e}_i,0)$. We can see that a lower bound on the value function of sensor $\ell$ can be obtained as a solution of the following minimization problem over $u_{\ell}^b$, where $u_{\ell}^b$ is the control action for sensor $\ell$ given a belief state $\emb{e}_b$
\begin{align}
J^{\ell}(\emb{e}_b) = \min_u \Bigg\{&\sum_{j=0}^{u-1}\sum_{i=1}^m [\emb{e}_bP^j]_i \lambda(i,\ell)T(i,\ell)+\sum_{i=1}^m [\emb{e}_b P^u]_i \lambda(i,\ell)T_0(i,\ell)\nonumber\\
&+ c\sum_{i=1}^m [\emb{e}_bP^{u+1}]_i+\sum_{i=1}^m [\emb{e}_bP^{u+1}]_i J^{\ell}(\emb{e}_i)\Bigg\}
\label{eq,final_lb_per_sensor}
\end{align}

Equation (\ref{eq,final_lb_per_sensor}) together with (\ref{eq,decomposed_valfunc}) define a lower bound on the total expected cost. To further tighten the bound we can now optimize over a matrix $\Lambda$ for every value of $c$, where $\Lambda(c)$ is an $m\times n$ matrix with the $(i,\ell)$ entry equal to $\lambda(i,\ell)$, i.e., $\Lambda(c)=\{\lambda(i,\ell)\}$. Hence,
\begin{align}
J(\emb{e}_b) = \max_{\Lambda(c)}\sum_{\ell = 1}^n\min_{u^b_{\ell}} \Bigg\{&\sum_{j=0}^{u-1}\sum_{i=1}^m [\emb{e}_bP^j]_i \lambda(i,\ell)T(i,\ell)+\sum_{i=1}^m [\emb{e}_b P^u]_i \lambda(i,\ell)T_0(i,\ell)\nonumber\\
&+ c\sum_{i=1}^m [\emb{e}_bP^{u+1}]_i+\sum_{i=1}^m [\emb{e}_bP^{u+1}]_i J^{\ell}(\emb{e}_i)\Bigg\}
\label{eq:final_lb}
\end{align}
\[
\mbox{subject to}~~~~ \Lambda \emb{1}_n = \emb{1}_m
\]
where $\emb{1}_m$ is a column vector of all ones of length $m$. A closed form solution for (\ref{eq:final_lb}) cannot be obtained, and hence, we solve for $J(\emb{e}_b)$ numerically. First, we fix $\Lambda$ and use policy iteration \cite{bertsekas07} to solve for the control of each sensor at each state. Then, we change $\Lambda$ and repeat the process. The envelope of the generated value functions (corresponding to different instants of $\Lambda$) is hence a lower bound on the optimal value function.

\section{Numerical Results} \label{sec:num_res}
In this section, we show some simulation results illustrating the performance of the policies we derived in previous sections.  These results will be for one-dimensional sensor networks, but the general behavior should extend to two-dimensional networks.  In each simulation run, the object was initially placed at the center of the network and the location of the object was made known to each sensor.  A simulation run concluded when the object left the network.  The results of many simulation runs were then averaged to compute an average tracking cost and an average energy cost.  To allow for easier interpretation of our results, we then normalized our costs by dividing by the {\em expected} time the object spends in the network.  We refer to these normalized costs as costs per unit time, even though the true costs per unit time would use the {\em actual} times the object spent in the network (the difference between the two was found to be small).

For the non-learning policies, the value of $T^\Delta(b,\ell)$ for each $b$ and $\ell$ was generated using 200 Monte Carlo simulations.  The results of 50 simulation runs were averaged when plotting the curves.  For the learning policies, the values for $T^\Delta$ were initialized to those obtained from the non-learning approach using greedy sensor selection as a baseline.  A constant step size of 0.01 was used in the learning algorithm.  First, 100 simulation runs were performed but the results were not recorded while the values for  $T^\Delta$ stabilized.  Then an additional 50 simulation runs were performed ($T^\Delta$ continued to be updated) and these results were averaged when plotting curves.  In the case of \QMDP{} learning policies, computation time was saved by performing policy iteration only after every fifth simulation run.

We first consider a simple network that we term Network~A.  This is a one-dimensional network with 41 possible object locations where the object moves with equal probability either one to the left or one to the right in each time step.  There is a sensor at each of the 41 object locations that makes (when awake) a binary observation that determines without error whether the object is at that location.  Hamming cost is used for the tracking cost.

\begin{figure}
\centering
\begin{tabular}{cc}
\epsfig{file=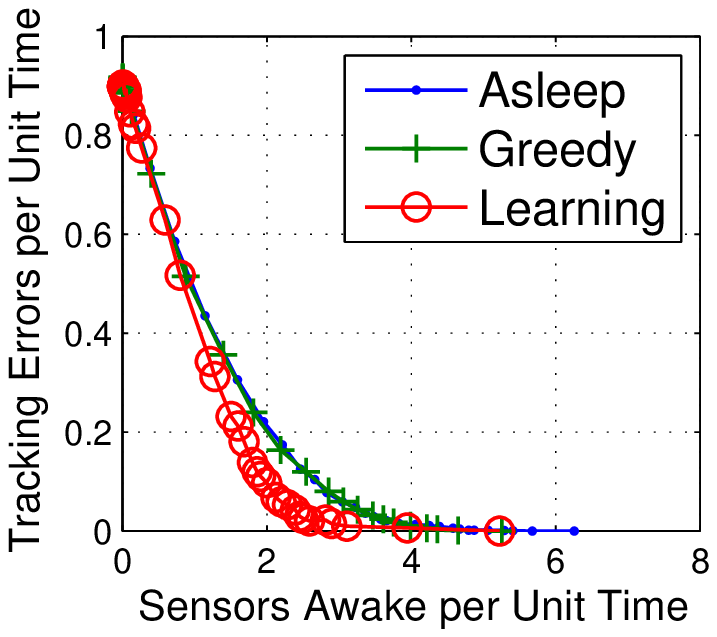,width=0.45\linewidth} &
\epsfig{file=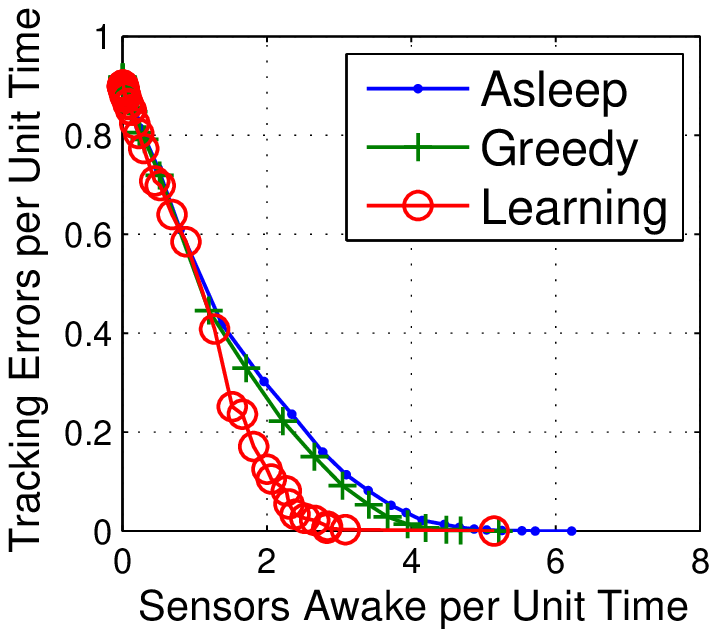,width=0.45\linewidth} \\
\mbox{(a)} & \mbox{(b)}
\end{tabular}
\caption{Tradeoff curves for Network~A: (a) \QMDP{} policies; (b) FCR policies}
\label{fig:hard41}
\end{figure}

For Network~A, we illustrate the performance of the \QMDP{} versions of our policies in Figure~\ref{fig:hard41}(a) and the FCR versions of our policies in Figure~\ref{fig:hard41}(b).

%
%\begin{figure}
%   \begin{center}
%      \includegraphics[width=3in]{hard41q_new.eps}
%      \caption{Tradeoff curves for \QMDP{} policies for Network~A. \label{fig:hard41q}}
%   \end{center}
%\end{figure}
%\begin{figure}
%   \begin{center}
%      \includegraphics[width=3in]{hard41f_new.eps}
%      \caption{Tradeoff curves for FCR policies for Network~A. \label{fig:hard41f}}
%   \end{center}
%\end{figure}
The curves labeled ``Asleep'' are for the nonlearning approach for computing $T^\Delta$ where we assume that all sensors are asleep as a baseline.  The curves labeled ``Greedy'' are for the nonlearning approach for computing $T^\Delta$ where we use a greedy algorithm to determine our baseline.  The curves labeled ``Learning'' employ our learning algorithm for computing $T^\Delta$.
%While we have included plots of cost per unit time in these figures for completeness, these plots do not convey much meaning and indeed can lead to confusion.  Consider that in the tradeoff curves of Figure~\ref{fig:hard41f} it appears that the Asleep policy results in the worst performance.  This is indeed the case.  However, in the plot of cost per unit time, it appears that the Asleep policy has the best performance.  The reason for this apparent contradiction is that we have substituted an approximate tracking cost for the actual tracking cost.  The result is that the original relationships between tracking cost and $c$ no longer hold when computing the sleeping policy.

From the tradeoff curves, it is apparent that using the learning algorithm to compute $T^\Delta$ results in improved performance.  A close inspection of Figures~\ref{fig:hard41}(a) and \ref{fig:hard41}(b) will reveal that the \QMDP{} policies perform somewhat better than their FCR counterparts.  This is consistent with what was observed in \cite{fuemmeler08}.

\begin{figure}
   \begin{center}
      \includegraphics[height=3.25in]{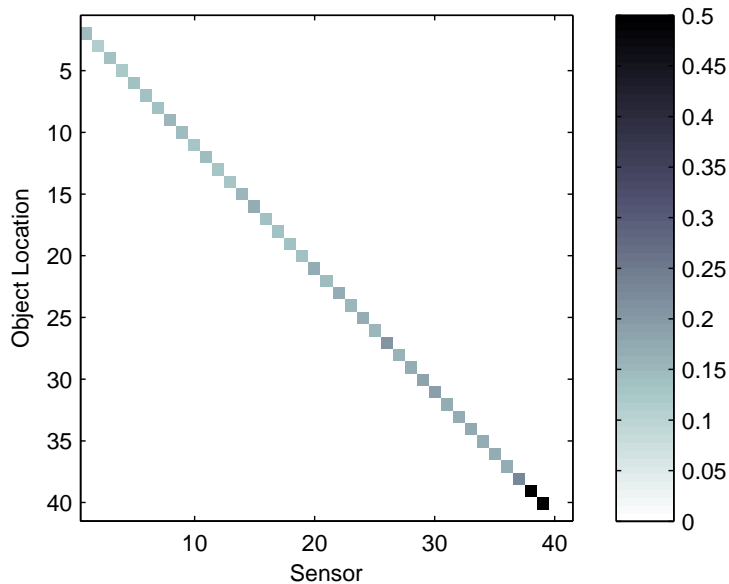}
      \caption{The final matrix for $T^\Delta$ for the \QMDP{} learning policy and small $c$ for Network~A. \label{fig:T41c0}}
   \end{center}
\end{figure}

\begin{figure}
   \begin{center}
      \includegraphics[height=3.25in]{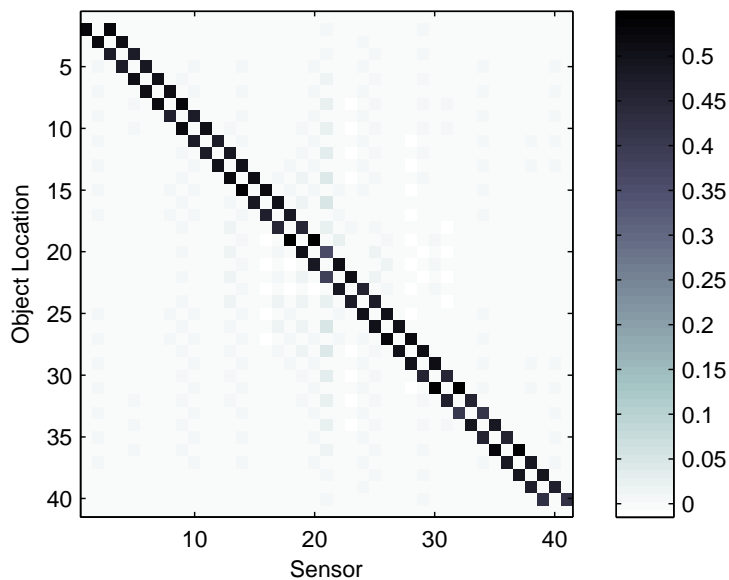}
      \caption{The final matrix for $T^\Delta$ for the \QMDP{} learning policy and large $c$ for Network~A. \label{fig:T41c1}}
   \end{center}
\end{figure}

It is instructive to consider the final matrix of values for $T^\Delta(b,\ell)$ that was obtained at the end of all learning algorithm simulations.  In Figures~\ref{fig:T41c0} and \ref{fig:T41c1} we plot this matrix for the \QMDP{} learning policy simulations for the smallest $c$ and for the largest $c$ used in simulation, respectively.
In Figure~\ref{fig:T41c0}, it is evident that only a single sensor has an impact for each value of $b$.  Due to the way our simulations worked, it is the sensor to the left that has the impact, but it could just as easily be the sensor to the right of the current object position. The fact that most of the nonzero values of the matrix are less than 0.5 reflects the fact that the sensor to the right of the current object location might wake up due to a sleep time selected at a previous time step.  In Figure~\ref{fig:T41c1}, it is evident that the sensors on either side of the current object location (which is actually not known since Figure~\ref{fig:T41c1} corresponds to the case where no sensors are awake) appear to have a major impact on the tracking cost.  There are nonzero values off the two main diagonals due to probabilistic nature of the learning process when the actual object location is not known.

\begin{table}
   \begin{center}
      \caption{Object movement for Network~B. \label{tab:moves}}
      \begin{tabular}{|l|r|r|r|r|}
         \hline
         {\em Change in Position} & 0 & 1 & 2 & 3  \\ \hline
         {\em Probability} & 0.3125 & 0.2344 & 0.0938 & 0.0156 \\ \hline
      \end{tabular}
   \end{center}
\end{table}

We now consider a new one-dimensional network termed Network~B.  The possible object locations are located on the integers from 1 to 21.  The object moves according to a random walk anywhere from three steps to the left to three steps to the right in each time step. The distribution of these movements is given in table \ref{tab:moves}. The change in position indicate movement by a corresponding number of steps to the right or to the left. There are 10 sensors in this network so that $m \neq n$. The locations of the sensors are given in Table~\ref{tab:sensloc} and awake sensors make Gaussian observations as in (\ref{eq:gaussian_obs}).
\begin{table}
   \begin{center}
      \caption{Sensor locations for Network~B. \label{tab:sensloc}}
      \begin{tabular}{|l|r|r|r|r|r|r|r|r|r|r|}
         \hline
         {\em Sensor} & 1 & 2 & 3 & 4 & 5 & 6 & 7 & 8 & 9 & 10 \\ \hline
         {\em Location} & 1.36 & 1.61 & 3.91 & 8.09 & 11.96 & 13.39 & 13.52 & 13.66 & 16.60 & 18.68 \\ \hline
      \end{tabular}
   \end{center}
\end{table}

\begin{figure}
\centering
\begin{tabular}{cc}
\epsfig{file=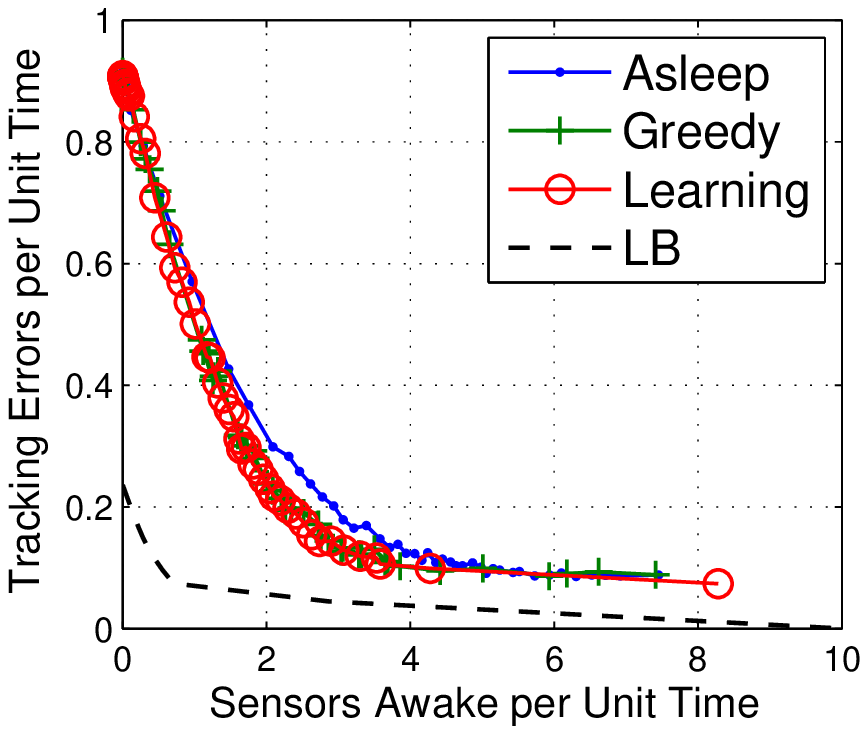,width=0.45\linewidth} &
\epsfig{file=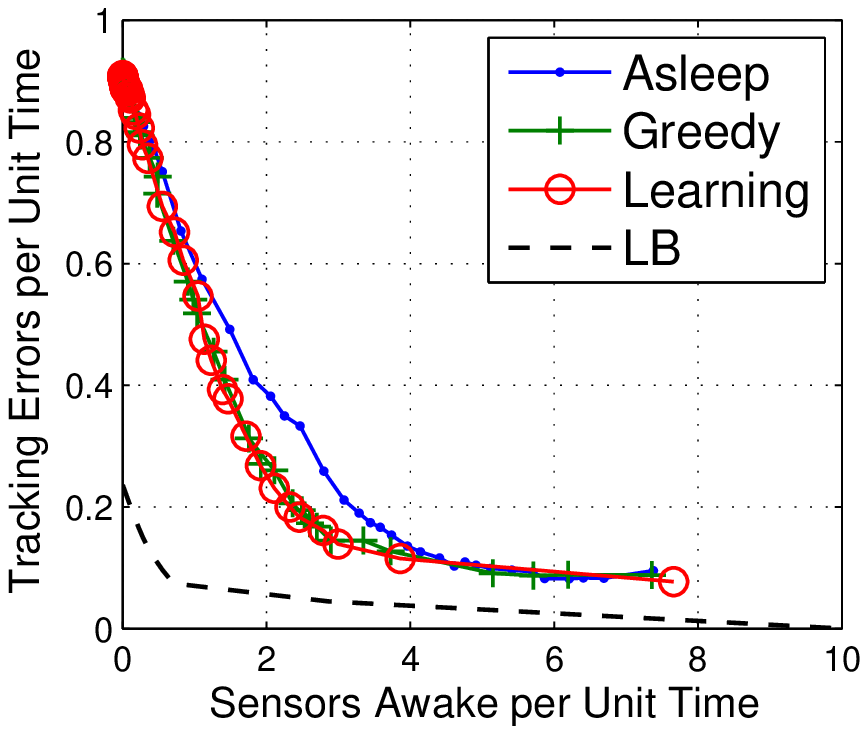,width=0.45\linewidth} \\
\mbox{(a)} & \mbox{(b)}
\end{tabular}
\caption{Tradeoff curves for Network~B and a lower bound: (a) \QMDP{} policies; (b) FCR policies}
\label{fig:hard21}
\end{figure}

\begin{figure}
   \begin{center}
      \includegraphics[height=3.25in]{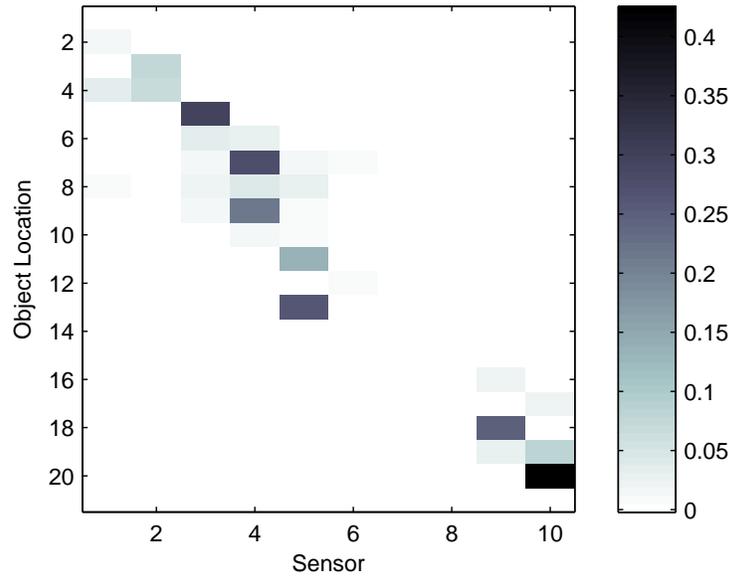}
      \caption{The final matrix for $T^\Delta$ for the \QMDP{} learning policy and small $c$ for Network~B. \label{fig:T21c0}}
   \end{center}
\end{figure}

\begin{figure}
   \begin{center}
      \includegraphics[height=3.25in]{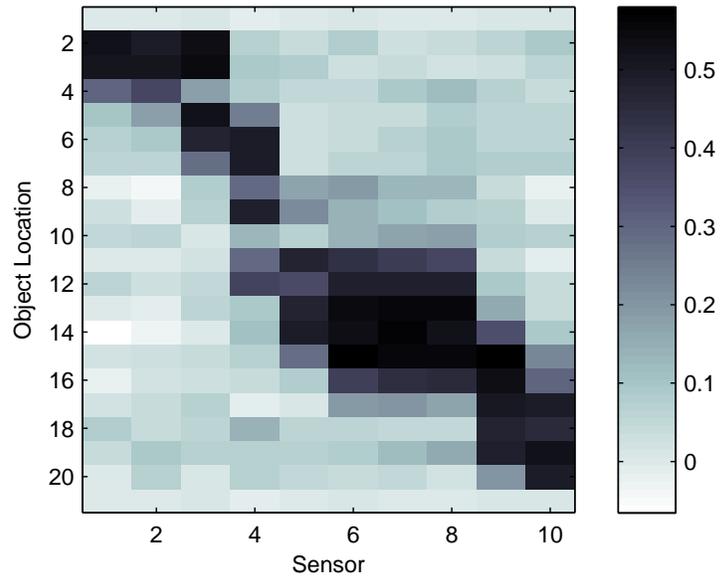}
      \caption{The final matrix for $T^\Delta$ for the \QMDP{} learning policy and large $c$ for Network~B. \label{fig:T21c1}}
   \end{center}
\end{figure}

Results for the \QMDP{} and FCR versions of our policies are shown in Figures~\ref{fig:hard21}(a) and \ref{fig:hard21}(b), respectively.
%
%\begin{figure}
%   \begin{center}
%      \includegraphics[width=3in]{hard21q_newest.eps}
%      \caption{Tradeoff curves for \QMDP{} policies for Network~B. \label{fig:hard21q}}
%   \end{center}
%\end{figure}
%\begin{figure}
%   \begin{center}
%      \includegraphics[width=3in]{hard21f_newest.eps}
%      \caption{Tradeoff curves for FCR policies for Network~B. \label{fig:hard21f}}
%   \end{center}
%\end{figure}
The results confirm the same general trends observed for Network~A. The figures also show our derived lower bound on the energy-tracking tradeoff using the approach described in Sec. \ref{sec:lower_bound}. Not surprisingly, the lower bound is particularly loose at the high tracking cost regime, yet the gap is reasonably small for the low tracking error region. This is expected since the lower bound uses an all-awake assumption to lower bound the contribution of each sensor to the tracking error. However, it is worth mentioning that we can exactly compute the saturation point for the optimal scheduling policy, which matches the saturation limit of the shown curves, since every policy has to eventually meet the all-asleep performance curve when the energy cost per sensor is high. At that point, all sensors are put to sleep and hence the target estimate can only be based on prior information. The small gap at the low tracking error regime combined with the aforementioned saturation effect highlight good performance for our sleeping policies. For illustration, we plot the matrix for $T^\Delta$ for the \QMDP{} learning policy simulations for the smallest $c$ and for the largest $c$ when the object moves according to a symmetric random walk in Figures~\ref{fig:T21c0} and \ref{fig:T21c1}, respectively.
Note the difference between the rows corresponding to object locations 7 and 8 in Figure~\ref{fig:T21c0}.  Examining the sensor locations, we see that sensor 4 is located at 8.09.  This sensor is useful for distinguishing between object locations 6 and 8 (for an initial object position of 7) but is of less value for distinguishing between object locations 7 and 9 (for an initial object position of 8).  This is evidenced in the figure as a large value for $T^\Delta(7,4)$ and a small value for $T^\Delta(8,4)$.

To demonstrate that our techniques can be applied to an object that moves on a continuum, we define a new network, Network~C.  This network is identical to Network~B except for two changes.  First, the object can take locations anywhere on the interval $[1, 21]$.  Second, the object moves according to Brownian motion with the change in position between time steps having a Gaussian distribution with mean zero and variance 1.  As mentioned earlier, only FCR policies can be generated for this type of network.  Values of $T^\Delta$ were computed for each integer-valued object location on $[1,21]$ and linear interpolation used to compute values of $T^\Delta$ for other object locations.  Since continuous distributions cannot be easily stored, particle filtering techniques were employed (e.g., see \cite{doucet01}).  The number of particles used was 512 and resampling was performed at each time step.  As is consistent with particle filtering, in generating the sleep times the computation of future probability distributions was approximated through Monte Carlo movement of the particles.  The number of simulation runs that were averaged for each data point was increased to 200 for these simulations.

\begin{figure}
   \begin{center}
      \includegraphics[height=3.25in]{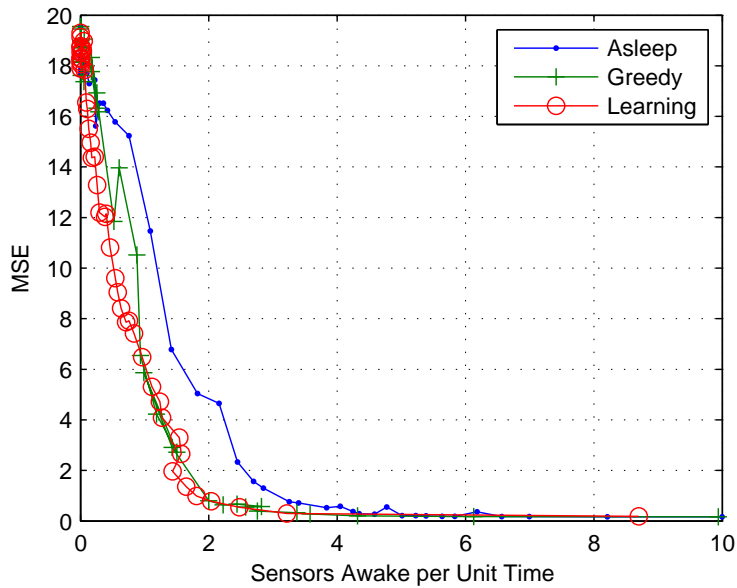}
      \caption{Tradeoff curves for FCR policies for Network~C. \label{fig:vhard21}}
   \end{center}
\end{figure}

Tradeoff curves for Network~C are shown in Figure~\ref{fig:vhard21}.
Although the tradeoff curves are less smooth than before, this figure illustrates performance trends similar to those already seen.  The reason the curves are not as smooth is that occasionally the particle filter would fail to keep track of the distribution with sufficient accuracy.  This would cause the network to lose track of the object and cause abnormally bad tracking for that simulation run.  These outliers were not removed when generating the tradeoff curves.  A recovery mechanism would need to be added to the sleeping policies to overcome this limitation of particle filters.

\section{Conclusion} \label{sec:concl}
In this paper, we considered energy-efficient tracking of an object moving through a network of wireless sensors. While an optimal solution could not be found, it was possible to design suboptimal, yet efficient, sleeping solutions for general motion, sensing, and cost models. We proposed \QMDP{} and FCR approximate policies, where in the former, the system is assumed to be perfectly observable after control, and in the latter, to be  totally unobservable. We combined these approximations with a decomposition of the optimization problem into simpler per-sensor subproblems, and developed learning and non-learning based approaches to compute the parameters of each subproblem. The learning-based \QMDP{} policies were shown to provide the best energy-tracking tradeoff. In the low tracking error regime, our sleeping policies approach a derived lower bound on the optimal energy-tracking tradeoff. %Furthermore, as the energy cost per sensor increases, looseness is reflected in the lower bound due to an all-awake assumption. At this regime, our sleeping policies reasonably saturate at the all-asleep operating point where only prior information becomes available.

Avenues for future research include developing distributed sleeping strategies in the absence of central control and solving the tracking problem for unknown or partially known object movement statistics.

%Distributed strategies for the scenario where a central controller is not available is an area for future research. Solving the tracking problem when the statistics for object movement are unknown or partially known presents another interesting challenge.
\bibliographystyle{IEEEtran}
% Put BibTeX in thesis.bib.
\bibliography{phdthesis}

\end{document}